\documentclass[twocolumn,showpacs,preprintnumbers,amsmath,amssymb,floatfix]{revtex4-1}

\usepackage{graphicx,psfrag}
\usepackage{dcolumn}
\usepackage{bm}
\usepackage{ulem}
\usepackage{amsmath}
\usepackage{setspace}
\usepackage{subfigure}
\usepackage{epsfig}

\usepackage[T1]{fontenc}
\usepackage[latin1]{inputenc}

\numberwithin{equation}{section}

\def\tens{\overleftrightarrow}

\begin{document}
\draft
\title{Supersymmetric Quantum Mechanics For Atomic Electronic Systems}

\author{Thomas Markovich}
\thanks{R.A. Welch Foundation Undergraduate Scholar, R.A. Welch Foundation Grant E-0608}
\affiliation{
Departments of Physics and Mathematics\\
University of Houston\\
Houston, TX 77204-5006
}
\author{Mason Biamonte}
\thanks{R.A. Welch Foundation Undergraduate Scholar, R.A. Welch Foundation Grant E-0608}
\affiliation{
Departments of Physics and Mathematic\\
University of Houston\\
Houston, TX 77204-5006
}

\author{Donald J. Kouri}
\thanks{Supported in part under R.A. Welch Foundation Grand E-0608}
\affiliation{
Departments of Chemistry, Mathematics,\\
Mechanical Engineering, and Physics\\
University of Houston\\
Houston, TX 77204-5006
}


\begin{abstract}

We employ our new approach to non-relativistic supersymmetric quantum mechanics (SUSY-QM), (J. Phys. Chem. A 114, 8202(2010)) for any number of dimensions and distinguishable particles, to treat the hydrogen atom in full three-dimensional detail. In contrast to the standard one-dimensional radial equation SUSY-QM treatment of the hydrogen atom, where the superpotential is a scalar, in a full three-dimensional treatment, it is a vector which applies regardless of the electron angular momentum. The original scalar Schr\"odinger Hamiltonian operator is factored into vector ``charge'' operators:  $\vec Q$ and $\vec Q^{\dagger}$.  Using these operators, the first sector Hamiltonian is written as $\mathcal{H}_{1} = \vec Q^{\dagger}\cdot \vec Q  + E_{0}^{(1)}$. The second sector Hamiltonian is a tensor given by  $\tens{\mathcal{H}}_{2} = \vec{Q}\vec{ Q}^{\dagger} + E_{0}^{(1)}\tens{1}$ and is isospectral  with $\mathcal{H}_{1}$.   The second sector ground state, $\vec\psi_{0}^{(2)}$, can be used to obtain the excited state wave functions of the first sector by application of the adjoint charge operator. Alternatively, $\vec{Q}$ applied to analytical, sector one excited states yield analytical results for the sector two vector eigenstates. Several of these are plotted for illustration. We then adapt the aufbau principle to show this approach can be applied to treat the helium atom.

\end{abstract}

\pacs{31.15.A-, 11.30.Pb}

\maketitle

\section{Introduction}\label{intro}

In a previous publication, we have provided a generalization of supersymmetric quantum mechanics (SUSY-QM) to treat any number of dimensions or particles with a focus on its usefulness as a computational tool for calculating accurate excited state energies and wave functions \cite{Kouri:2010es}. We note that Stedman has presented a similar treatment for multi-dimensional systems. However, he generates extra sector states not present in our approach\cite{Stedman:1985tp}. His result is more complicated and the meaning of the sector Hamiltonians is not discussed. Stedman presents his equations for the hydrogen atom but does not solve any of the higher sector equations. Because of the significant analytical and computational ramifications, we here apply our multi-dimensional generalization of SUSY-QM to the hydrogen atom in full three-dimensional detail. This is of interest because, until now, the standard application of SUSY-QM to the hydrogen atom required that we first separate out the angular degrees of freedom -- effectively reducing the problem to a one-dimensional treatment \cite{Kirchberg:2003uo, Tangerman:1993uo, Lahiri:1987vg}. With our vector superpotential approach, one can deal with the full three-dimensional nature of the hydrogen atom.

Our approach provides, for the first time, a SUSY-QM framework that can be employed to treat non-hydrogenic atoms. For example, the standard SUSY-QM treatment of the hydrogen atom cannot be readily extended to the helium atom because it is impossible reduce it to a one-dimensional system. In addition, the form of the three-dimensional vector superpotential for the hydrogen atom is of interest in its own right. It is quite different from the radial superpotential obtained in earlier SUSY-QM studies of the hydrogen atom. The present study thus lays the groundwork for a systematic SUSY-QM study of excited state energies and wave functions of atoms. As we have discussed in our earlier studies, there appears to be a significant increase in the accuracy of both excited state energies and wave functions when one computes the sector two ground state energy and wave functions, followed by application of the adjoint ``charge operator'' to generate the sector one excited state wave functions \cite{ Kouri:2009bw, Kouri:2009kb, Kouri:2010es}. 

This paper is organized as follows. In section II, to make this paper self contained, we give a detailed description of the SUSY-QM formulation valid for any number of dimensions and particles. In section III, we describe the application of the theory to the hydrogen atom in three spatial dimensions. We give exact analytical expressions for the charge operators and obtain the ground state wave functions and energies of the sector two hydrogen atom tensor Hamiltonian. In fact, we note that since the exact solutions of the sector one hydrogen atom Hamiltonian are analytically known, we can also generate any of the excited states of the sector two problem. In Section IV, we outline how the approach can be applied to the helium atom. In Section V, we examine the ramifications of the aufbau principle in the second sector. Finally, in Section VI, we present our conclusions and plans for future research.

\section{Introduction to the Partner Hamiltonian Formulation of SUSY-QM in N-Dimensions}

The standard approach to supersymmetric quantum mechanics (SUSY-QM) provides an elegant scheme for solving one-dimensional problems, but until recently, it has not been generalized to multiple dimensions \cite{IntroSUSYCooper, Witten1,Cahill:1999vq}. We were originally attracted to SUSY-QM as a novel computational approach, since it converts a standard second order differential equation to one of first order. This reduction of order is achieved by factoring the Schr\"odinger Hamiltonian operator in terms of so-called ``charge'' operators, $Q$ and $Q^{\dagger}$. The simplest, and best known example is the one-dimensional harmonic oscillator where the  $Q$ and $Q^{\dagger}$ are the well-known lowering and raising operators \cite{dirac}.  We have solved a number of one-dimensional problems using this approach. Because of the success of our one-dimensional studies, we were motivated to generalize SUSY-QM to any number of dimensions or particles. 

Most previous attempts to generalize SUSY-QM to treat more than one spatial dimension and more than one particle generally have involved introducing additional ``spin-like''  degrees of freedom  \cite{Andrianov:1984sf,Das:1996sw,0305-4470-35-6-305, Andrianov:1986rm,Andrianov:1988yg,Andrianov:2002gf,A.A.Andrianov:1984qv,Eides:1984rz,Andrianov:1985ty,Andrianov:1984nr}. In our method \cite{Kouri:2010es}, we make use of a vectorial approach that simultaneously treats more than one dimension and any number of distinguishable particles (see also \cite{Stedman:1985tp}).  We consider, therefore, a system of $n$-particles in three-dimensional space. We denote the coordinates of particle $i$ by $(x_{i},y_{i}, z_i)$.  We then define an orthogonal hyperspace of dimension $3n$. We take the Hamiltonian for this system to be given by
\begin{equation}
\mathcal{H}_{1} = -\nabla^{2} + V_{1}
\end{equation}
where
\begin{equation}
\vec\nabla = \sum_{j} \vec\epsilon_{j}\frac{\partial}{\partial u_{j}}
\end{equation}
and $\vec\epsilon_{j}\cdot\vec\epsilon_{k} = \delta_{jk}$. The subscript ``1'' indicates this is the ``sector one'' Hamiltonian. For simplicity we take the masses of the particles to be equal and use units such that $\hbar^{2}/2m = 1$. For the development here we assume a Cartesian coordinate space, but have provided an extension to more general curvilinear coordinates in a previous publication \cite{Kouri:2010es}.

As per usual in quantum mechanics, the ground-state wave function is a solution of the Schr\"odinger equation,
\begin{equation}
\mathcal{H}_{1}\psi_{0}^{(1)} = E_{0}^{(1)}\psi_{0}^{(1)}.
\end{equation}
We also emphasize that the lowest energy state, $\psi_{0}^{(1)} $,  is nodeless.

We now define a vector superpotential, $\vec W$, as
\begin{equation}
\vec W= -\vec\nabla \ln \psi_{0}^{(1)},
\end{equation}
which is to say
\begin{equation}
\vec W =\sum_{j=1}^{3n}\vec\epsilon_{j} W_{j} = -\sum_{j=1}^{3n}\vec\epsilon_{j} \frac{\partial}{\partial u_{j}}\ln\psi_{0}^{(1)}.
\end{equation}
It is straightforward to see that one can write $\mathcal{H}_{1}$ in terms of $\vec W$ as
\begin{equation}
\begin{split}
(\mathcal{H}_{1}-E_{0}^{(1)} )= (-\vec\nabla + \vec W)\cdot (+\vec\nabla + \vec W) \\ = (-\partial_{i}+W_{i}) (\partial_{i}+W_{i}),
\end{split}
\end{equation}
where, according to the Einstein convention, we sum over repeated indices.
Since $(\vec\nabla + \vec {W})\psi_{0}^{(1)} \equiv \vec0$,
it is clear that $(\mathcal{H}_{1}-E_{0}^{(1)} )\psi_{0}^{(1)} =0 $ as required.
The SUSY charge operators are also vectors with components defined by:
\begin{equation}
Q_{i} = \partial_{i} + W_{i},\;  Q_{i}^{\dagger } = -\partial_{i} + W_{i}.
\end{equation}

We can now define the sector two Hamiltonian such that, above the ground-state $(E_{0}^{(1)})$, it is isospectral with $\mathcal{H}_{1}$.
We do this as follows, for the first excited state in sector one we can write
\begin{equation}\label{eqn19}
Q_{i}^{\dagger} \cdot Q_{i} \psi_{1}^{(1)} = (E_{1}^{(1)} - E_{0}^{(1)})\psi_{1}^{(1)}.
\end{equation}
We then form the tensor product by operating on the left with $\vec Q_{1}$ so that
\begin{equation}
(\vec Q\vec Q^{\dagger})\cdot \vec Q \psi_{1}^{(1)} =  (E_{1}^{(1)} - E_{0}^{(1)}) \vec Q_{1}\psi_{1}^{(1)}.
\end{equation}
That is to say, using Einstein notation, 
\begin{equation}
(Q_{i}Q_{j}^{\dagger}) Q_{j} \psi_{1}^{(1)} =  (E_{1}^{(1)} - E_{0}^{(1)}) Q_{i}\psi_{1}^{(1)}.
\end{equation}
It then follows that $\vec Q\psi_{1}^{(1)}$ is an eigenstate of the tensor Hamiltonian
$\tens{\mathcal{H}}_{2} = (\vec Q\vec Q^{\dagger})$
with energy $E_{0}^{(2)} = E_{1}^{(1)} -  E_{0}^{(1)}$.  Since we are free to set the
energy origin, taking $E_{0}^{(1)} =0$ gives $E_{0}^{(2)} = E_{1}^{(1)}$.
It is also clear that $\vec Q\psi_{0}^{(1)}$ cannot generate a lower energy
eigenstate of $\tens{\mathcal{H}}_{2}$ since  $\vec Q\psi_{0}^{(1)} = \vec 0$,
so that $\vec Q\psi_{1}^{(1)}$ is indeed proportional to the ground state of $\tens{\mathcal{H}}_{2}$. The precise relation between the two sets of states is
given by
\begin{equation}
\psi_{n+1}^{(1)} = \frac{1}{\sqrt{E_{n+1}^{(1)}-E_{0}^{(1)}}}\vec Q^{\dagger}\cdot \vec\psi_{n}^{(2)}.
\end{equation}

\section{SUSY-QM For The three-dimensional Hydrogen Atom}
\begin{figure*}
\centering
\mbox{\subfigure{\includegraphics[width=0.33333\linewidth]{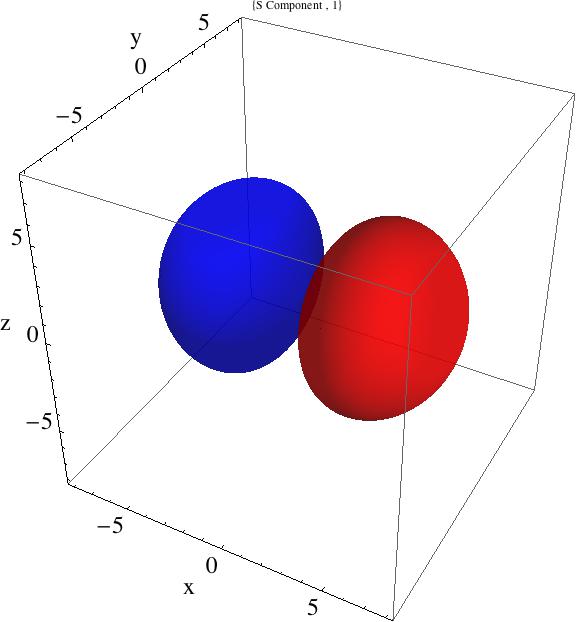}}\quad
\subfigure{\includegraphics[width=0.33333\linewidth]{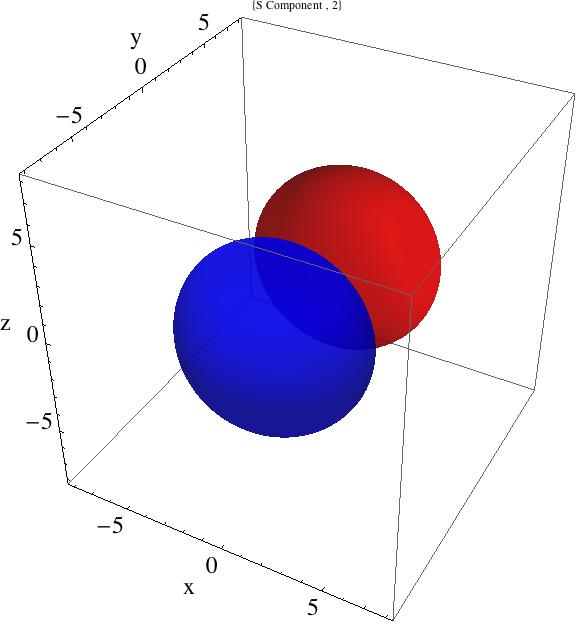} }\quad
\subfigure{\includegraphics[width=0.33333\linewidth]{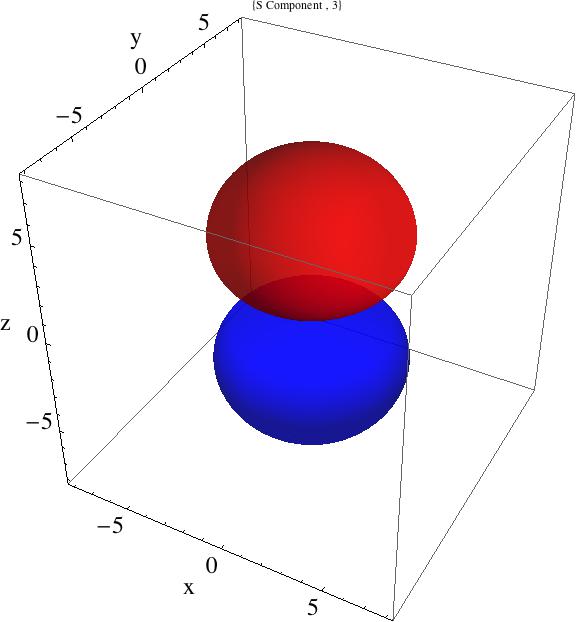} }
}
\caption{\label{s} The three components of the wave function for $\vec{\psi}_{1,S}^{(2)} $. Here, blue corresponds to positive values and red to negative. }
\end{figure*}

\begin{figure*}
\centering
\mbox{\subfigure{\includegraphics[width=0.33333\linewidth]{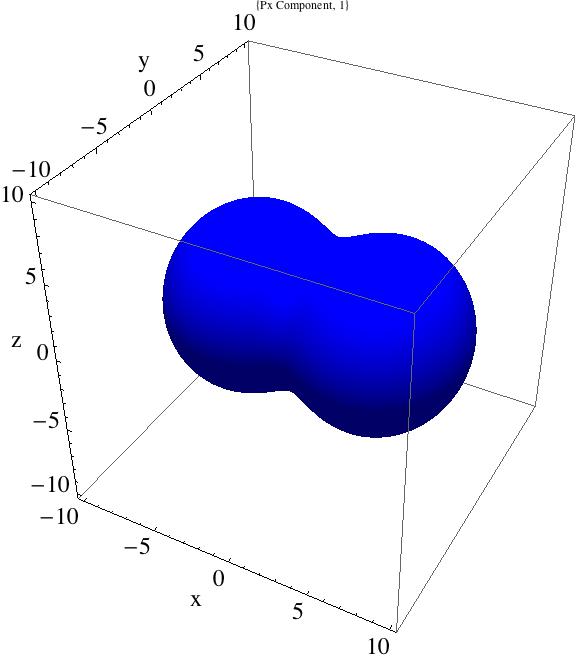}}\quad
\subfigure{\includegraphics[width=0.33333\linewidth]{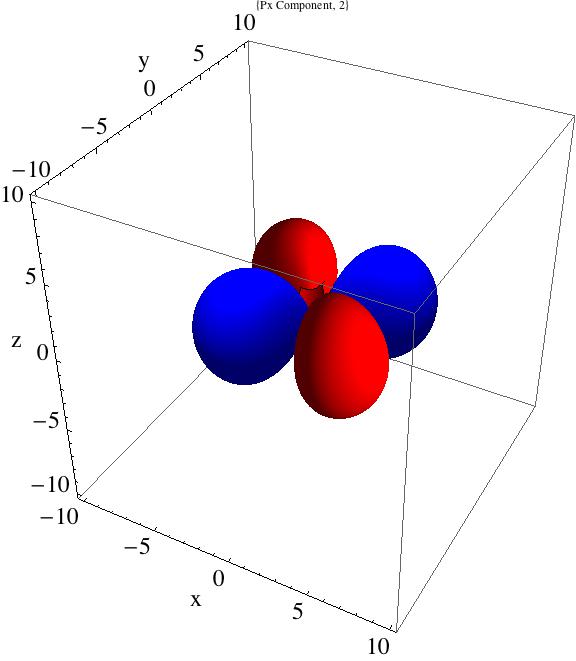} }\quad
\subfigure{\includegraphics[width=0.33333\linewidth]{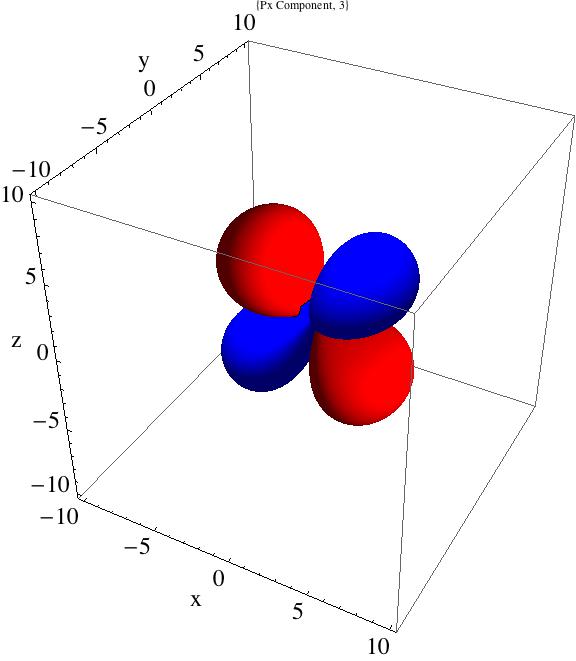} }
}
\caption{ \label{px} The three components of the wave function for $\vec{\psi}_{1,p_x}^{(2)}$. Here, blue corresponds to positive values and red to negative. }
\end{figure*}

We now consider the hydrogen atom. We begin by noting that the ground state is exactly given by
\begin{equation}
\psi_{1,0,0} = \frac{e^{-r}}{\sqrt{\pi}},
\end{equation}
where we have set the Bohr radius equal to 1. The Hamiltonian (in atomic units) is simply
\begin{equation}\label{neqn13}
\mathcal{H} = -\frac{1}{2} \nabla^2 - \frac{1}{r}.
\end{equation}
Then, the vector superpotential is given by
\begin{equation}\label{neqn13}
\vec{W} = -\nabla \ln { \psi_{1,0,0}} = \hat{r},
\end{equation}
where $\hat{r}$ is a unit vector in the direction of $\vec{r}$. This is an extremely interesting result. First, we see that the superpotential for the Coulomb interaction is, itself, non-singular. Second, in the standard approach, because the angular degrees of freedom have already been separated out, the superpotential is a scalar and it depends on the angular momentum squared ({\it i.e.} on $l(l+1)$). The precise form for the ground state ($l=0$) is
\begin{equation}\label{neqn13}
W_{radial} = 1
\end{equation}
In three dimensions, we have
\begin{equation}\label{neqn13}
\vec{W} = \vec{\epsilon_x} \frac{x}{r} + \vec{\epsilon_y} \frac{y}{r}  + \vec{\epsilon_z} \frac{z}{r} = \hat{r}.
\end{equation}
The magnitude of $\vec{W}$ is equal to the radial superpotential, as one expects, but the individual components are radically different. Note that these components can also be written solely in terms of angular functions (the direction cosines of $\vec{r}$). To obtain the atomic potential for hydrogen, we form
\begin{equation}\label{neqn13}
\vec{W} \cdot \vec{W} - \nabla \cdot \vec{W} = 1-\left ( \frac{3}{r}  -  \frac{ x^2 + y^2 + z^2 }{r^3} \right)
\end{equation}
\begin{equation}\label{7}
= 1 - \frac{2}{r} = -2 E_0 - \frac{2}{r_0}.
\end{equation}
Now we recall that
\begin{equation}\label{neqn13}
H \psi_{m_l} = E_n \psi_{m_l}
\end{equation}
and
\begin{equation}\label{neqn13}
-\frac{1}{2} \nabla^2 \psi_{m_l} = \left [ E_n + \frac{1}{r} \right] \psi_{m_l}
\end{equation}
yields
\begin{equation}\label{10}
\nabla^2 \psi_{1,0,0} = \left [2 E_0 + \frac{2}{r} \right] \psi_{1,0,0}.
\end{equation}
Since the ground state energy of hydrogen in atomic units is -1/2, we find that Equations \eqref{7} and \eqref{10} are consistent and we have obtained the correct vector superpotential. Of great interest is the wave equation for the sector two problem. This Hamiltonian is given by 
\begin{equation}\label{neqn13}
\tens{\mathcal{H}}_2 = -\frac{1}{2} \nabla \nabla + \frac{1}{2} \left[  \vec{W} \vec{W} + \nabla \vec{W}  \right].
\end{equation}
In the case of the hydrogen atom, because we have exact analytical expressions for the excited states of $\mathcal{H}_1$, it is a simple matter to generate analytical expressions for all the states of the sector two Hamiltonian. It is convenient to label the sector two states with an index indicating the $n^{th}$ energy state (i.e., we use the principle quantum number n = 1,2,...) along with the quantum numbers of the sector one excited state from which they are obtained. Thus, the four degenerate ground states of $\tens{\mathcal{H}}_2$ will be denoted by $\vec{\psi}^{(2)}_{1,2p_x}$, $\vec{\psi}^{(2)}_{1,2p_y}$, $\vec{\psi}^{(2)}_{1,2p_z}$, $\vec{\psi}^{(2)}_{1,2s}$. We choose here to use the real states rather than those labeled by $m_l = \pm1$ and $m_l = 0$ values. We find that these solutions are given by 
\begin{equation}\label{neqn13}
\vec{\psi}^{(2)}_{1,2p_x} = N \left[  \hat{i} e^{-r/2} + \frac{x \hat{r}}{2} e^{-r/2}  \right],
\end{equation}
\begin{equation}\label{neqn13}
\vec{\psi}^{(2)}_{1,2p_y} = N \left[  \hat{j} e^{-r/2} + \frac{y \hat{r}}{2} e^{-r/2}  \right],
\end{equation}
\begin{equation}\label{neqn13}
\vec{\psi}^{(2)}_{1,2p_z} = N \left[  \hat{k} e^{-r/2} + \frac{z \hat{r}}{2} e^{-r/2}  \right],
\end{equation}
\begin{equation}\label{neqn13}
\begin{split}
\vec{\psi}^{(2)}_{1,2s} = -N \frac{\vec{r}}{2} e^{-r/2}.
\end{split}
\end{equation}
These equations can be verified by simply applying $\vec{Q}$ to the first excited state wave functions of sector one. It is also easily verified that $\vec{Q}^{\dagger}$ acting on these states regenerates the $\psi^{(1)}_{2p}$ and $\psi^{(1)}_{2s}$ states. Furthermore, in Figures \ref{s} and \ref{px}, we provide plots of the $\vec{\psi}^{(2)}_{1,2s}$ and $\vec{\psi}^{(2)}_{1,2p_x}$. It is straight forward to see that $\vec{\psi}^{(2)}_{1,2p_y}$ and $\vec{\psi}^{(2)}_{1,2p_z}$ are both similar to $\vec{\psi}^{(2)}_{1,2p_x}$.

\section{An Approximate superpotential for the Helium Atom}
It is of interest to begin exploring how our approach to multidimensional SUSY-QM would deal with a two electron atom. It is clear that the usual radial (one-dimensional) hydrogen atom SUSY-QM treatment is not readily generalizable to deal with helium. We have carried out a Quantum Monte Carlo study of the sector one ground state of helium using the Pad\`e Jastrow trial wave function:
\begin{equation}
\psi^{(1)}_{T,\alpha} = e^{-2 r_1} e^{-2 r_2} e^{ \frac{r_{12}}{ 2(1+\alpha r_{12}) } },
\end{equation}
with the optimum $\alpha$ given by $\alpha = 0.353$. This yields an energy of $E^{(1)}_1 \approx$ 2.878, which is in error by about 1$\%$. This error is reasonable for a simple treatment neglecting relativistic interactions. The approximate $\vec{W}$ is generated from
\begin{equation}
\vec{W}(\vec{r_1},\vec{r_2}) = - \vec{\nabla} \ln \psi^{(1)}_{T,\alpha}
\end{equation}
\begin{equation}
= -\vec{\nabla} \left [  -2 r_1 - 2 r_2 + \frac{r_{12}}{  2 (1+\alpha r_{12} ) }  \right].
\end{equation}
Here,
\begin{equation}
\vec{\nabla} = \hat{\epsilon_{1x}} \frac{\partial}{\partial x_1} + \hat{\epsilon_{1y}} \frac{\partial}{\partial y_1} + \hat{\epsilon_{1z}} \frac{\partial}{\partial z_1} + 
\hat{\epsilon_{2x}} \frac{\partial}{\partial x_2} + \hat{\epsilon_{2y}} \frac{\partial}{\partial y_2} + \hat{\epsilon_{2z}} \frac{\partial}{\partial z_2},
\end{equation} 
where the $\{ \hat{\epsilon}_{ij} \}$ are orthonormal vectors. The resulting vector superpotential for the Pad\`e-Jastrow trial function is readily found to be
\begin{equation}
\vec{W}_{He(PJ)} = 2 \hat{r}_1 + 2 \hat{r}_2 - \hat{r}_{12} \left[ 1 - \frac{\alpha}{ \left ( 1 + \alpha r_{12} \right ) } \right ].
\end{equation}
Thus, the structure of $\vec{W}_{He(PJ)} $ is analogous to $\vec{W}_H$ in that Coulomb interactions generate vector superpotentials that involve unit vectors anti-parallel to the direction of the forces. This is true in general for Coulombic interactions. This emphasizes the important distinction between our three-dimensional SUSY-QM treatment of an atom and the standard hydrogen atom one-dimensional radial SUSY-QM. We are currently carrying out calculations of the sector two ground state for helium using the above $\vec{W}_{He(PJ)}$. There are several possible approaches to be explored. One is to follow our earlier two-dimensional study and employ a Rayleigh-Ritz variational method\cite{Kouri:2010es}. This is immediately applicable to the helium problem and should work without difficulty. However, it will involve much more computational effort since helium is a six-dimensional system as opposed to the two-dimensional systems studied earlier. The second approach we intend to explore is the Dirac-Frankel-McLachlan time-dependent variational method\cite{Raab2000674}. The results of these studies will be reported later. 

\section{Aufbau Approach For Excited States}
For multielectron atoms, it becomes necessary to consider how the aufbau principle acts in the second sector to permit efficient calculations of sector one excited states. This is because we can use this principle to design reasonable trial wave functions for a variational approach to the sector two ground state. In this section, we consider a simple aufbau description of the sector one helium excited states in order to design an approximate sector two ground state of helium. We assume that in the first excited state of sector one, we have one electron in the $1\it{S}$ orbital, given by wave function $\alpha$, and one electron in the $2\it{S}$ orbital, given by wave function $\beta$ where
\begin{equation}
\alpha(r) = \frac{e^{-2r}}{\sqrt{\pi }}
\end{equation}
and
\begin{equation}
\beta(r) = \frac{e^{-r}}{4\sqrt{2\pi }}(1-r).
\end{equation}
Then, it is of interest to take the product of these states such that we have $\alpha(r_1) \beta(r_2)$, to which we can apply our $\vec{Q}$ to find 
\begin{equation}\label{ab12}
\vec{Q} \left( \alpha(r_1) \beta(r_2) \right) = -e^{-2r_1 - r_2} \left [ 2 \hat{r}_1 \left ( 1 - r_2 \right) + \hat{r}_2 \right] \equiv \vec{\phi}_1^{(2)}.
\end{equation}
It is clear that to find the second sector state associated with $\alpha(r_2) \beta(r_1)$, we simply need to interchange labels 1 and 2 in Equation \eqref{ab12} to get
\begin{equation}
\vec{Q} \left( \alpha(r_2) \beta(r_1) \right) = -e^{-2r_2 - r_1} \left [ 2 \hat{r}_2 \left ( 1 - r_1 \right) + \hat{r}_2 \right] \equiv P_{12} \vec{\phi}_1^{(2)},
\end{equation}
where $P_{12}$ exchanges the electron labels. Then, we can use this as a ``building block'' to construct our ground state in the second sector in the following way. Suppose we want to create the the spin triplet state, we find that we need to subtract the first building block from the second. This gives us a second sector result of 
\begin{equation}
\begin{split}
\vec{\psi}_{1,triplet}^{(2)} =& -e^{-2r_1 - r_2} \left [ 2 \hat{r}_1 \left ( 1 - r_2 \right) + \hat{r}_2 \right] +\\
 &e^{-2r_2 - r_1} \left [ 2 \hat{r}_2 \left ( 1 - r_1 \right) + \hat{r}_1 \right]. 
\end{split}
\end{equation}
And similarly, we can find the second sector electronic structure singlet state by simply adding the two building blocks as given below.
\begin{equation}
\begin{split}
\vec{\psi}_{1,singlet}^{(2)} =& -e^{-2r_1 - r_2} \left [ 2 \hat{r}_1 \left ( 1 - r_2 \right) + \hat{r}_2 \right] -\\
 &e^{-2r_2 - r_1} \left [ 2 \hat{r}_2 \left ( 1 - r_1 \right) + \hat{r}_1 \right]. 
\end{split}
\end{equation}
Indeed, by taking the scalar product with $\vec{Q}^{\dagger}$, we can verify that $\vec{\psi}_{1,triplet}^{(2)}$ and $\vec{\psi}_{1,singlet}^{(2)}$ give the appropriate spatial wave functions. This is to say that, to within a multiplicative constant, we get that 
\begin{equation}
\vec{Q}^{\dagger} \cdot \vec{\psi}_{1,triplet}^{(2)} = \psi^{(1)}_1 = \alpha(r_1) \beta(r_2) - \alpha(r_2) \beta(r_1)
\end{equation}
and 
\begin{equation}
\vec{Q}^{\dagger} \cdot \vec{\psi}_{1,singlet}^{(2)} = \psi^{(1)}_2 = \alpha(r_1) \beta(r_2) + \alpha(r_2) \beta(r_1).
\end{equation}
From this, we observe that the aufbau principle in the second sector is remarkably simple. We merely need to take the building block $\vec{\phi}_1^{(2)}$ and antisymmetrize or symmetrize appropriately. One interesting thing to note is that our ``building block'', $\vec{\phi_1^{(2)}}$, is neither symmetric nor antisymmetric under particle exchange. 

However, this basis doesn't include the correlation. To do this, we can multiply our antisymmetrized second sector wave function by a correlation function, given by the Pad\'e-Jastrow function which only depends on $r_{12}$. It is clear, then, that because our correlation function is only a function of $r_{12}$, its symmetry will not be affected by the application of $\vec{Q}$ and, thus, we can simply multiply it by our second sector state of interest (where the minus corresponds to the triplet and the plus to the singlet):
\begin{equation}
\begin{split}
 \vec{\psi}_{1,triplet}^{(2)} =& e^{\frac{r_{12}}{2(1+\delta r_{12})}} \left ( -e^{-2r_1 - r_2} \left [ 2 \hat{r}_1 \left ( 1 - r_2 \right) + \hat{r}_2 \right] \mp \right.\\
 & \left. e^{-2r_2 - r_1} \left [ 2 \hat{r}_2 \left ( 1 - r_1 \right) + \hat{r}_1 \right] \right). 
 \end{split}
\end{equation}
These equations appear to be qualitatively correct but one will insert variational parameters ({\it e.g.}, effective charges, {\it etc.}) when doing computations. The next step is to perform numerical calculations using $\vec{\psi}^{(2)}_{1,triplet}$ in the second sector to find energies generated only from the assumptions above and compare the results with standard approaches. We are currently exploring more complex atomic structures to generalize this further. 

\section{Conclusions}
In this paper we have shown how our multi-dimensional generalization of SUSY-QM can be applied to the hydrogen atom. Previously, most detailed attempts to treat the hydrogen atom first separated the angular degrees of freedom, leaving a one-dimensional radial wave equation. It was then possible to obtain the SUSY-QM factorization, yielding a scalar superpotential that, for the $l=0$ states, is simply $W=1$. While these results are interesting, the one-dimensional radial SUSY-QM approach is not readily generalizable to treat even the helium atom.

In our approach, the full three-dimensional character of the hydrogen atom is considered, with the result being a vector-valued superpotential, $\vec{W}$, which for the hydrogen atom, is $\vec{W} = \hat{r}$. That is, the vector superpotential points in the opposite direction of the attractive Coulomb force between the electron and the nucleus. This is interesting also because, although the Coulomb potential is singular, its vector superpotential is not. It is important to note that such a superpotential was also obtained earlier by Stedman \cite{Stedman:1985tp}. However, his sector two Hamiltonian differs from ours and produces ``extra'' states that are not degenerate with sector one.

The fact that $\vec{W}$ for the three-dimensional hydrogen atom is a vector does not, in any way, modify the sector one dynamical equation. However, the sector two situation is radically affected! In the one-dimensional SUSY-QM case, there is no significant change in the basic mathematical structure of the sector two partner Hamiltonian. In the multi-dimensional case, the sector two Hamiltonian is a tensor. However, we have shown in previous studies, that many of the standard computational techniques remain valid. Of particular interest is the Dirac-Frankel-McLachlan Variational Method, since this is known to deal better with higher-dimensional systems\cite{Raab2000674}.

In the case of the hydrogen atom, it is straight forward to generate all the sector two eigenstates. This is a consequence of the fact that exact analytical eigenstates of the three-dimensional hydrogen atom are known. It is then easy to apply the charge operator, $\vec{Q}$, to the excited hydrogen atom states and obtain sector two eigenstates. (We note that because of the four fold degeneracy for the sector two ground state, the resulting eigenstates can be super-posed in any manner convenient for the study at hand). It is of considerable interest to begin exploring how our multi-dimensional SUSY-QM treatment can be applied to the helium atom. In this case, the exact sector one ground state is, of course, unavailable. In our previous one and two-dimensional studies we have considered other systems for which an exact $\vec{W}$ was not possible. In the case of helium, we chose to examine an accurate Pad\`e-Jastrow approximation to the sector one ground state. In this case, it is easy to obtain an analytical (albeit approximate) $\vec{W}$ that displays very reasonable intuitive character. In direct analogy with the exact hydrogen atom $\vec{W}$, we find that the $\vec{W}_{He(PJ)}$ vector superpotential consists of a combination of unit vectors that again, are anti-parallel to the Coulomb forces associated with the helium atom potential energy. The next step in our study will consist of computations of a sector two ground state, which will allow us to obtain an approximate helium atom sector one first excited state energy and wave function.

Future studies will explore extending the approach to more than two electron atoms. There, the issue will be taking account of the electrons' spin degrees of freedom. Our current plan is to employ the ``spin-free'' techniques of Matsen \cite{Matsen:1970ig, Matsen:1964gt, Matsen:1966ig, Matsen:1966cq, Matsen:1968hx, Matsen:1969jd, Matsen:1969ch, Matsen:1969dn, Matsen:1971dh, Matsen:1971ko, Matsen:1971bc} and others\cite{pauncz}.

We have also generalized the aufbau principle to work in the second sector Hamiltonian, demonstrating that we are able to produce reasonable forms of excited states by simply using hydrogenic orbitals. The equations have a reasonable structure but variational computations are necessary. We shall report these results later.

\bibstyle{apsrev}
\bibliography{papersBib}

\end{document}